\pdfoutput=1
\PassOptionsToPackage{obeyspaces}{url}
\documentclass[twocolumn,prb,a4paper,final,superscriptaddress,longbibliography,nofootinbib,floatfix]{revtex4-1}

\usepackage{amsmath,amssymb}
\usepackage{graphicx}
\usepackage{comment}
\usepackage{bm}
\usepackage[colorlinks=true,allcolors=blue]{hyperref}
\usepackage{color}
\usepackage{subfigure}
\usepackage{CJKutf8}
\usepackage{ulem}
\usepackage{placeins}
\usepackage{cancel}

\newcommand{\Eprint}[2]{\href{#1}{\urlstyle{same}\nolinkurl{#2}}}

\usepackage{soul}
\setstcolor{red}

\newcommand{\pd}{\partial}

\newcommand{\sign}{\qopname\relax o{sgn}}

\newcommand{\const}{\text{const}}

\newcommand{\E}{\qopname\relax o{E}}

\newcommand{\calH}{{\cal H}}

\newcommand{\n}{{\bf n}}
\newcommand{\D}{{\bf D}}

\newcommand{\tilx}{\tilde{x}}
\newcommand{\tily}{{\tilde{y}}}
\newcommand{\tilth}{\tilde{\vartheta}}

\newcommand{\DW}{{\mathrm{DW}}}
\newcommand{\SW}{{\mathrm{SW}}}
\newcommand{\LW}{{\mathrm{LW}}}

\begin{document}
\title{Skyrmion crystal phase on a magnetic domain wall in chiral magnets}
	
\date{\today}
\author{Yuki Amari}
\affiliation{Research and Education Center for Natural Sciences, Keio University, 4-1-1 Hiyoshi, Yokohama, Kanagawa 223-8521, Japan}
\affiliation{Department of Physics, Keio University, 4-1-1 Hiyoshi, Yokohama, Kanagawa 223-8521, Japan}

\author{Muneto Nitta}
\affiliation{Department of Physics, Keio University, 4-1-1 Hiyoshi, Yokohama, Kanagawa 223-8521, Japan}
\affiliation{Research and Education Center for Natural Sciences, Keio University, 4-1-1 Hiyoshi, Yokohama, Kanagawa 223-8521, Japan}
\affiliation{International Institute for Sustainability with Knotted Chiral Meta Matter (SKCM$^2$), Hiroshima University, 1-3-2 Kagamiyama, Higashi-Hiroshima, Hiroshima 739-8511, Japan}

\vspace{.5in}
	
\begin{abstract}
We study a magnetic domain wall in the ferromagnetic phase 
in chiral magnets in two dimensions 
with an in-plane easy-axis anisotropy 
and an out-of-plane Zeeman magnetic field, 
and find a 
chiral soliton lattice (spiral) phase
beside a ferromagnetic phase 
inside the domain line, 
where the former 
represents a domain-wall skyrmion crystal 
from the bulk point of view.
We first determine the phase diagram on the domain wall by numerically constructing domain-wall solutions.
We then analytically reproduce the phase diagram in a domain-wall theory
(a chiral double sine-Gordon model) that we construct within the moduli approximation by treating the Zeeman magnetic field perturbatively. 
While we find good agreements between the phase diagrams of the numerical and effective theory methods, the numerical solution exhibits a decomposition of the topological charge into a bimeron which cannot be captured by the effective theory.

\end{abstract} 
\maketitle

\newpage

\section{Introduction}
\label{sec:intro}
Skyrmions are topological solitons first introduced by Skyrme as a model of nuclei \cite{Skyrme:1962vh} 
and are later identified with the 
baryons in 
a certain
limit of 
quantum chromodynamics (QCD) \cite{Witten:1983tx}. 
As the two dimensional analog of nuclear skyrmions,  
magnetic skyrmions 
\cite{Bogdanov:1989,Bogdanov:1995} 
 have attracted much attention in chiral magnets which are stabilized by a Dzyaloshinskii-Moriya (DM) interaction \cite{Dzyaloshinskii,Moriya:1960zz}.
Such magnetic skyrmions have been observed in laboratory experiments \cite{doi:10.1126/science.1166767,doi:10.1038/nphys2045,doi:10.1038/nature09124} and have applications for information carriers in magnetic storage devices \cite{Nagaosa2013}. 
Both isolated skyrmions \cite{doi:10.1126/science.1240573} and skyrmion tubes \cite{Wolf_2021} have been also observed. 
See Ref.~\cite{Nagaosa2013} 
for a review of magnetic skyrmions.

On the other hand, 
magnetic domain walls are other solitonic objects that appear in chiral magnets with an easy-axis potential.
They have also been a subject of particular study due to their application to magnetic memories \cite{doi:10.1126/science.1145799,KUMAR20221}.
Apart from skyrmions and 
domain walls,  
a lot of studies have been devoted to 
various topological objects 
such as 
monopoles \cite{tanigaki2015,fujishiro2019topological}, 
Hopfions \cite{Sutcliffe:2018vcb} 
and 
instantons \cite{Hongo:2019nfr}
(see Ref.~\cite{GOBEL20211} for a review).

Chiral magnets have a quite rich phase structure governed by 
skyrmions and domain walls (solitons). 
The most conventional state realized for small DM interaction 
is a uniform state, 
the ferromagnetic phase. 
There, both the skyrmions and domain walls can appear as positive energy solitons. 
There are typically two kinds of inhomogeneous ground states made of 
topological solitons 
due to the presence of the DM interaction. 
One is a spiral phase, which can be seen as a chiral soliton lattice (CSL)
\cite{togawa2012chiral,KISHINE20151,PhysRevB.97.184303}, 
where the energy of a single soliton 
or domain wall is negative 
and thus one dimensional modulated states 
composed of solitons or domain walls have a lower energy than 
uniform states. 
The other is the skyrmion lattice (crystal) phase where 
the energy of a single skyrmion is negative \cite{doi:10.1126/science.1166767,doi:10.1038/nature09124,doi:10.1038/nphys2045,Rossler:2006,Lin:2014ada,Han:2010by,Ross:2020hsw}.

As mentioned, 
magnetic skyrmions
are important ingredients 
for applications to information carriers in magnetic storage devices \cite{Nagaosa2013}.
In such a case, 
skyrmions should be manipulated 
by applying an electric current;  
the skyrmions start to move along the electric current. 
However, 
the trajectories of the skyrmions 
are bent because of the skyrmion Hall effect 
\cite{zang2011dynamics,Chen:2017,Jiang:2017,litzius2017skyrmion}.
This is one of the major difficulties when controlling the motion of skyrmions 
for such a technology.
To overcome this difficulty, 
skyrmions absorbed into a domain wall can be considered 
in which case 
skyrmions can move only along the domain wall. 
Such composite objects of skyrmions and a domain wall are called domain-wall skyrmions\footnote{
The term 
``domain-wall skyrmions'' was first introduced in Ref.~\cite{Eto:2005cc} 
in which Yang-Mills instantons 
in the bulk are 
3D skyrmions inside a domain wall. 
The terminology of this paper is different 
from Ref.~\cite{Eto:2005cc}; what was studied there should be called 
domain-wall instantons 
in the current terminology.
}
and are expected to be useful for constructing easily controllable future magnetic memories. 
Domain-wall skyrmions have been studied in 
quantum field theory \cite{Nitta:2012xq,Kobayashi:2013ju} (see also Refs.~\cite{Jennings:2013aea,Bychkov:2016cwc})  
and more recently both theoretically \cite{PhysRevB.99.184412,Kim:2017lsi,Lee:2022rxi,PhysRevB.102.094402,KBRBSK,Ross:2022vsa,Amari:2023gqv,PhysRevB.109.014404,Gudnason:2024shv,Lee:2024lge,Leask:2024dlo} and experimentally \cite{Nagase:2020imn,li2021magnetic,Yang:2021} in chiral magnets.\footnote{ 
These are the two-dimensional counterparts of 
the three-dimensional domain-wall skyrmions in quantum field theory~\cite{Nitta:2012wi,Nitta:2012rq,
Gudnason:2014nba,Gudnason:2014hsa,Eto:2015uqa,Nitta:2022ahj}, with recent interests in application to QCD in a strong magnetic field 
\cite{Eto:2023lyo,Eto:2023wul} 
or rapid rotation~\cite{Eto:2023tuu}.
} 
 In particular, for an {\it out-of-plane} easy-axis 
anisotropy, the worldline of a domain-wall skyrmion is bent to form a cusp at the position of the skyrmion~\cite{PhysRevB.99.184412,KBRBSK,Amari:2023gqv}.
 In our previous paper \cite{Amari:2023bmx}, 
such a cusp was studied using both the analytic method of the moduli approximation,  
sometimes called the Manton approximation 
\cite{Manton:1981mp,Eto:2006pg,Eto:2006uw} 
(a double sine-Gordon equation in Ref.~\cite{PhysRevB.99.184412})  
and a numerical simulation.
In the spiral (CSL) phase,
a domain-wall skyrmion is unstable to decay into a bimeron through a reconnection process \cite{Amari:2023bmx}  
(see also Ref.~\cite{PhysRevB.106.224428} 
for a bimeron in the case of Zeeman magnetic field). 
In addition to this 
a chain of domain-wall skyrmions 
can be stable in the spiral phase \cite{Amari:2023bmx}.\footnote{A three-dimensional version of a domain-wall skyrmion chain has been recently found in QCD \cite{Eto:2023wul}.
}

In this paper, we study domain-wall skyrmions 
in chiral magnets in two dimensions 
with an 
{\it in-plane}
easy-axis anisotropy, 
instead of the {\it out-of-plane} easy-axis 
anisotropy studied in 
the above mentioned works.
We also apply 
an out-of-plane Zeeman magnetic field\footnote{Dynamics of single domain-wall skyrmions in the same setting has been studied in Ref.~\cite{Jiwen:2024}. In addition, Ref.~\cite{Gong:2017} has experimentally and theoretically studied spin structures on a domain wall in chiral magnets that are equipped with both in-plane and out-of-plane easy-axis anisotropies. }.
We note that the direction of a single straight domain wall 
is not arbitrary and there is an energetically preferable direction, unlike the case 
of the out-of-plane  
easy-axis anisotropy for which 
the direction of a domain wall is arbitrary.

First we perform numerical simulations 
to construct domain-wall configurations, 
and determine a phase diagram of the lowest energy single
domain-wall states, 
that is a phase diagram of the ground states inside a single domain wall.
We find that 
the ground state of the domain wall 
is not only in the uniform ferromagnetic phase (unlike the case 
of the out-of-plane easy-axis anisotropy) 
but also in a CSL or spiral phase, even though the bulk is in the ferromagnetic phase 
but not in the spiral (CSL) phase.
Such a CSL on the domain wall is nothing but 
a skyrmion crystal along the domain-line from the bulk point of view. 
In this case, the domain wall is straight without any cusps, 
in contrast to the out-of-plane easy anisotropy 
in which case a domain wall has a cusp at the point of a skyrmion as mentioned above.
Such a skyrmion crystal phase of the domain wall appears in a stripe region of the phase diagram.
We also observe in a certain parameter region inside the skyrmion crystal phase that the topological skyrmion charge is decomposed into 
two peaks, and each carries a half topological charge 
and can be regarded as a meron.

Second, we study such a skyrmion crystal state 
analytically.
To this end, we 
construct 
the effective theory on a single domain wall 
by using the moduli approximation 
\cite{Manton:1981mp,Eto:2006pg,Eto:2006uw}, 
to yield a chiral double sine-Gordon model.\footnote{
A chiral double sine-Gordon model in the bulk appears 
in chiral magnets with the both easy-axis anisotropy 
and a Zeeman magnetic field, e.~g.~\cite{Ross:2020orc} and also in QCD 
\cite{Eto:2021gyy,Eto:2023tuu,Eto:2023rzd}.
}
We then find that a CSL state appears in a quite close 
region with the domain-wall skyrmion crystal phase region of the 
numerically determined phase diagram.
We compare configurations of 
the CSL on the domain wall with the domain-wall skyrmion crystal obtained by numerical simulations 
and find qualitative agreements between them.
However, the moduli approximation cannot capture the decomposition of topological charge into two 
merons.
Although the domain wall has the energetically preferable direction, 
we can rotate it with applying magnetic fields alternately,  
resulting in a wavy domain-wall skyrmion crystal. 
This can be seen both from the effective theory and numerical simulations. 


This paper is organized as follows. In Sec.~\ref{sec:model}, we introduce the model of a chiral magnet and review some properties of its domain-wall solutions. 
In Sec.~\ref{sec:chain-numerics}, we numerically construct a skyrmion crystal on a single domain wall, i.e., domain-wall skyrmion crystal, and map out a phase diagram showing the domain-wall skyrmion crystal appears as the lowest energy single domain wall state.
In Sec.~\ref{sec:moduli_approx}, we give an analysis of domain-wall skyrmion crystals using the moduli approximation.
We construct domain-wall skyrmion crystals in the effective theory on a domain wall and compare those with the numerical solutions. 
Finally, we summarize and discuss some open questions in Sec.~\ref{sec:discussion}.

\section{Model and domain-wall configuration}
\label{sec:model}

In this section, we introduce the Hamiltonian we study in the paper and describe its domain-wall solutions. We also discuss properties of the domain wall that we will utilize in the proceeding sections.

\subsection{Model}
\label{subsec:model}

Low energy phenomena in chiral magnets with the ferromagnetic exchange interaction can be described by a continuum theory for a three-component unit vector $\n=(n_x,n_y,n_z)$ representing the normalized magnetization vector, i.e., $\n =\mathbf{M}/M_s$ with the magnetization vector $\mathbf{M}$ and saturation magnetization $M_s$.
The continuum theory that we study in this paper is defined by a Hamiltonian of the form
\begin{equation}
    \begin{split}
        \calH=&\frac{1}{2} \partial_{k} \n \cdot \partial_{k} \n+\D_{k} \cdot\left(\n \times \partial_{k} \n\right)
\\
&\qquad\qquad\qquad
-\mathbf{H}\cdot \n + \frac{\mu^{2}}{2}\left(1-n_{x}^{2}\right)
\label{eq:bulk_theory}
    \end{split}
\end{equation}
with $k=x, y$, where the energy scale is $|J|M_s^2$ with the exchange constant $J$ positive.
The first term is the $O(3)$ nonlinear sigma model, which is equivalent to the continuum version of the Heisenberg exchange interaction. The second term is the DM interaction, and the third is the Zeeman interaction where ${\bf H} = \mathbf{B}/JM_s$ with the external magnetic field $\mathbf{B}$.  The last term is an in-plane easy-axis anisotropy with a positive coupling constant $\mu$.
We consider the magnetic field applied perpendicular to the $x\text{-}y$ plane, $\mathbf{H}=(0,0,-h)$ and employ the Rashba-Dresselhaus type DM vectors
\begin{align}
\begin{aligned}
& \D_{x} =-\kappa(\cos \vartheta,-\sin \vartheta, 0) \\
& \D_{y} =-\kappa(\sin \vartheta, \cos \vartheta, 0)
\end{aligned}
\end{align}
where $\kappa$ and $\vartheta$ are constants.
The spin-spiral period giving typical length scale is defined as $L_D=2\pi/|\kappa|$.

The topological charge density for skyrmions is defined by
\begin{equation}
    {\cal Q}=\frac{\varepsilon^{kl}}{8\pi} \n \cdot (\partial_{k}\n \times \partial_{l}\n ) \ .
\end{equation}
The integral of ${\cal Q}$ over the $x$-$y$ plane provides an integer counting the number of skyrmions.

We introduce new coordinates by rotating the $x$-$y$ plane with an angle $\nu$ as
\begin{align}
\left(\begin{array}{l}
\tilde{x} \\
\tilde{y}
\end{array}\right)=\left(\begin{array}{cc}
\cos \nu & -\sin \nu \\
\sin \nu & \cos \nu
\end{array}\right)\left(\begin{array}{l}
x \\
y
\end{array}\right) \ .
\end{align}
In terms of the new coordinates, the DM interaction can be rewritten as
\begin{align}
\D_{k} \cdot\left(\n \times \partial_{k} \n\right)
=\tilde{\D}_{k} \cdot\left(\n \times \tilde{\partial}_{k} \n\right)
\end{align}
where $\left(\tilde{\partial}_{x}, \tilde{\partial}_{y}\right)=\left(\partial_{\tilde{x}}, \partial_{\tilde{y}}\right)$ and
\begin{align}
\begin{aligned}
& \tilde{\D}_{x}=-\kappa(\cos \tilde{\vartheta},-\sin \tilde{\vartheta}, 0) \\
& \tilde{\D}_{y}=-\kappa(\sin \tilde{\vartheta}, \cos \tilde{\vartheta}, 0)
\end{aligned}
\end{align}
with $\tilde{\vartheta} \equiv \vartheta-\nu$.
The other terms in the Hamiltonian are invariant under the spatial rotation.
For simplicity, we choose the $\tilde{x}$ axis as the direction parallel to domain walls.

\subsection{domain-wall solutions}
\label{subsec:domain-wall}

\begin{figure}[t]
    \centering
    \includegraphics[width=\linewidth]{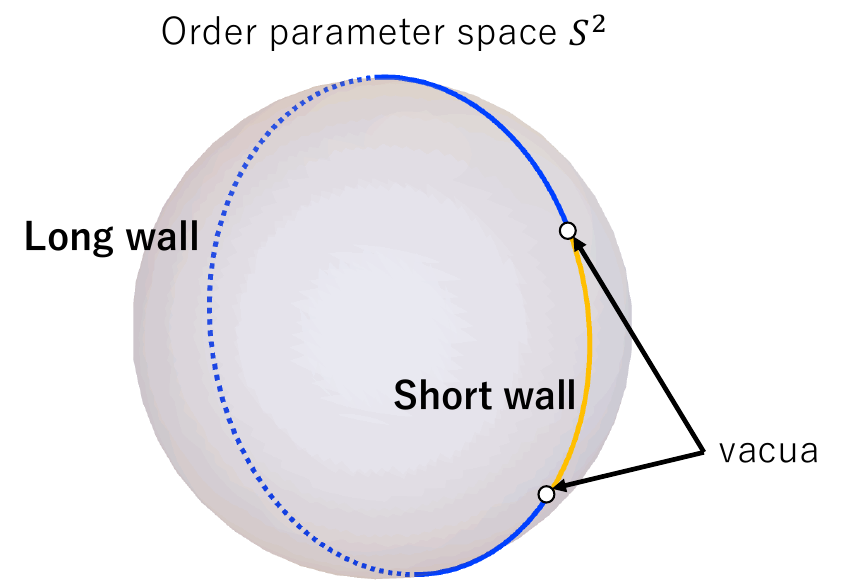}
    \caption{Arcs corresponding to short walls and long walls on the OPS $S^2$. The yellow curve represents the arc corresponding to a short wall, and the blue one corresponds to a long wall.}
    \label{fig:OPS}
\end{figure}

Domain-wall solutions parallel to the $\tilde{x}$ axis can be derived using an ansatz of the form
\begin{equation}
    \n =(\cos f, \sin f \cos \phi, \sin f \sin \phi)
    \label{eq:n_parametrization}
\end{equation}
with 
\begin{equation}
    f=f(\tilde{y}), \qquad \phi= \const. \ .
    \label{eq:ansatz}
\end{equation}
Substituting this ansatz into the Hamiltonian, we obtain
\begin{align}
    \calH
    &=\frac{1}{2}\left(\partial_{\tilde{y}} f\right)^{2}
    +\kappa \cos \tilde{\vartheta} \sin \phi ~\partial_{\tilde{y}} f
    \notag\\
    &\qquad\qquad
    +h\sin \phi \sin f
    +\frac{\mu^{2}}{2} \sin ^{2} f \ .
\label{eq:wall-energy}
\end{align}
This is called the chiral double sine-Gordon model
, which is the double sine-Gordon model with the topological term proportional to $\kappa$. Domain-wall solutions in this model have been studied in Ref.~\onlinecite{Ross:2020orc}.
For later convenience, we define
\begin{equation}
\begin{split}
     & \gamma = \frac{|h|}{\mu^2}, 
     ~~ s_1 =\sign(h), ~~ s_2 =\sign(\kappa \cos\tilde{\vartheta}).
\end{split}
\end{equation}
Note that in this paper, we let $\gamma \in[0,1)$ where the Hamiltonian has two vacua given by
\begin{equation}
\begin{split}
     &\phi=s_3 \frac{\pi}{2}, \\
     &\sin f = - s_1s_3 \gamma  
\end{split}
\end{equation}
with $s_3=\pm1$, which leads to
\begin{equation}
    {\mathbf n} =\left(\pm \sqrt{1-\gamma^2}, 0, -s_1\gamma \right) \ .
\end{equation}
For later convenience, we name the vacua as 
\begin{align}
    &\n_{\rm vac_A} \equiv \left(-s_1s_2 \sqrt{1-\gamma^2}, 0, -s_1\gamma \right) \ ,
    \\
    &\n_{\rm vac_B} \equiv\left(s_1s_2 \sqrt{1-\gamma^2}, 0, -s_1\gamma \right) ,
\end{align}
and explicitly refer to $\n_{\rm vac_A}$ as the vacuum A and $\n_{\rm vac_B}$ as the vacuum B.
When the Hamiltonian has two vacua, we have two types of domain walls according to the way to connect the vacua in the order parameter space (OPS) $S^2$. They are called short walls and long walls. In Fig.~\ref{fig:OPS}, we show a schematic picture of arcs representing short walls and long walls on $S^2$. On the other hand, if the Hamiltonian has only one vacuum, there exists only one type of domain wall (soliton) corresponding to a large circle on $S^2$, starting and ending at the vacuum.

\begin{figure}[t]
    \centering
    \includegraphics[width=\linewidth]{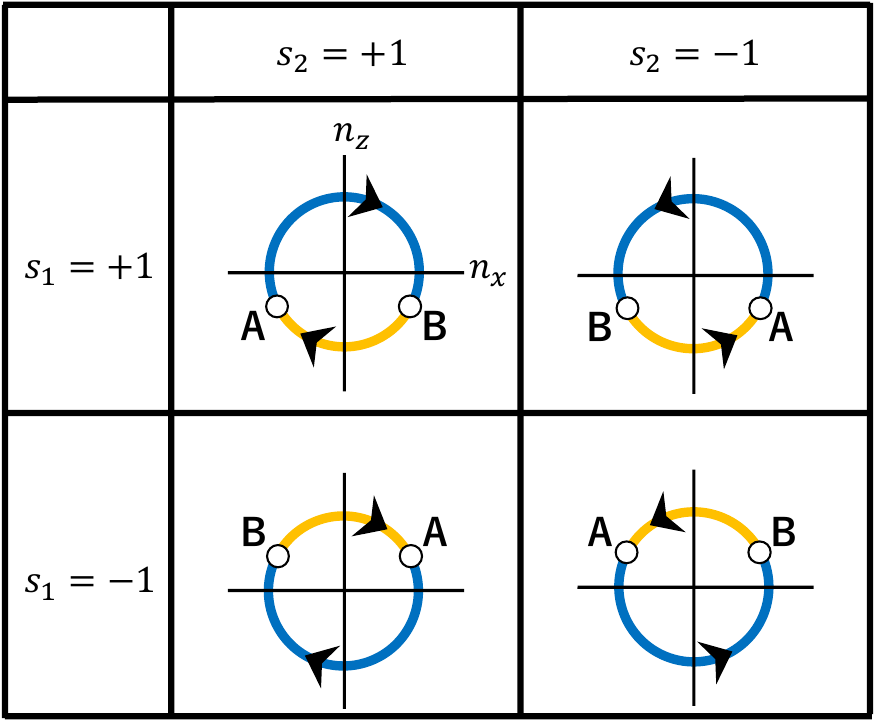}
    \caption{Classification of stable domain walls. Domain walls corresponding to the yellow arcs are short walls $\SW^-$, and blue arcs are $\LW^-$. 
    The quantity $s_1$ determines the position of the vacua that are represented by white dots on the circles on the $n_x\text{-}n_z$ plane. 
    The field value of stable walls should change in the direction represented by the black arrows determined by $s_2$.
    $\SW^-$ begins from vacuum B and ends at vacuum A, and conversely, $\LW^-$ begins from vacuum A and ends at vacuum B. One can also consider $\SW^+$ and $\LW^+$ for each $s_1$ and $s_2$. Their field values change in the directions opposite to the black arrows. It means that $\SW^+$ ($\LW^+$) begins from the vacuum A (B) and ends at the vacuum B (A).  
     }
    \label{fig:wall_classification}
\end{figure}

The equations of motion with respect to $f$ and $\phi$ are respectively given by 
\begin{align}
    &\pd^2_{\tily}f-h\sin\phi\cos f-\frac{\mu^2}{2}\sin(2f)=0 \, , 
    \label{eq:eomDW_f}
    \\
    &\cos\phi~(\kappa \cos\tilth \pd_\tily f +h \sin f )=0 \, .
    \label{eq:eomDW_phi}
\end{align}
Eq.~\eqref{eq:eomDW_f} is equivalent to the double sine-Gordon equation.
Eq.~\eqref{eq:eomDW_phi} is clearly satisfied if $\phi = \pm \pi/2$.
Using it, one obtains, from Eq.~\eqref{eq:eomDW_f}, a first-order ordinary differential equation for $f$ of the form
\begin{equation}
    \pd_\tily f =\pm\sqrt{\mu^2\sin^2 f +2 s_1s_3 |h|\sin f +C}
    \label{eq:ODE_f}
\end{equation}
where $C$ is an integration constant.
We impose the boundary conditions that $f$ decays into one vacuum at $\tily=-\infty$: 
\begin{equation}
    \lim_{\tily\to -\infty}\sin f  = -s_1s_3 \gamma, ~~
    \lim_{\tily\to-\infty}\pd_\tily f =0 \ .
\end{equation}
Then, one obtains $C=h^2/(2\mu^2)$ and it enables us to write Eq.~\eqref{eq:ODE_f} as
\begin{equation}
    \pd_\tily f =\pm |\mu||\sin f + s_1s_3\gamma| \,.
    \label{eq:ODE_f_simplified}
\end{equation}
In this case, short walls satisfy $s_1s_3\cdot\sin f \leq -\gamma$, and long walls satisfy $s_1s_3\cdot\sin f \geq -\gamma$. 
In addition, we have a further classification of the domain-wall solutions associated with the sign in Eq.~\eqref{eq:ODE_f_simplified}.
We denote the lower-energy  (higher-energy) short walls $\SW^-$ ($\SW^+$), and similarly the lower-energy  (higher-energy) long walls $\LW^-$ ($\LW^+$). 
Note that the sign does not necessarily coincide with that in Eq.~\eqref{eq:ODE_f_simplified}.
In Fig.~\ref{fig:wall_classification}, we show a schematic picture for the classification of arcs on the $n_x\text{-}n_z$ plane corresponding to $\SW^-$ and $\LW^-$.

\begin{figure*}[t]
    \centering
    \includegraphics[width=0.8\linewidth]{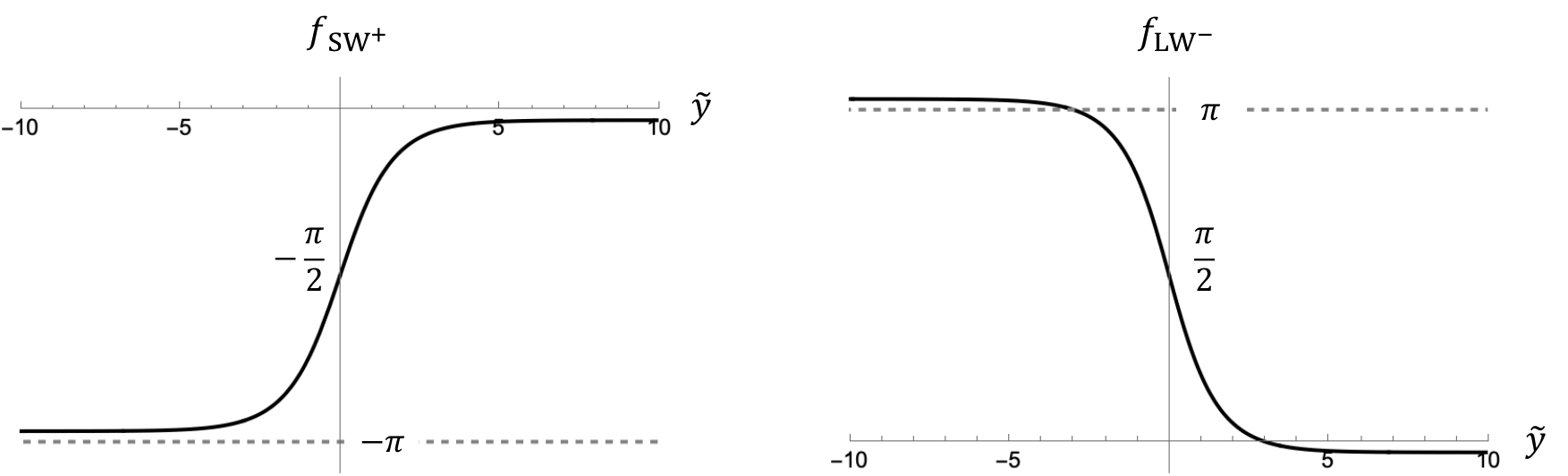}
    \caption{Profile of the domain-wall solutions. The left figure shows the profile of $f_{\SW^+}$, and the right one shows that of $f_{\LW^-}$. We have used $\mu=1.0, ~|h|=0.1,$ and $ s_1=s_2=s_3=+1$.}
    \label{fig:wall_profile}
\end{figure*}

The domain-wall solutions can be given by the form
\begin{widetext}
    \begin{align}
    f_{\SW^\pm} &= \pm s_2s_3\cdot 2\arctan\left[
    \sqrt{\frac{1-\gamma}{1+\gamma}}\tanh\left(\frac{\sqrt{1-\gamma^2}}{2}(\mu\tily-\chi) \right)
    \right]
    -s_1s_3 \frac{\pi}{2}
    \label{eq:sol_SW}
    \\
    f_{\LW^\pm} &= \pm s_2s_3\cdot 2\arctan\left[
    \sqrt{\frac{1+\gamma}{1-\gamma}}\tanh\left(\frac{\sqrt{1-\gamma^2}}{2}(\mu\tily-\chi) \right)
    \right]
    +s_1s_3\frac{\pi}{2} \,.
    \label{eq:sol_LW}
\end{align}
where $\chi$ is a moduli parameter associated with the domain-wall position.
Note that both $f_{\SW^+}$ and $f_{\LW^-}$ satisfy the boundary conditions 
\begin{equation}
    \lim_{\tilde{y}\to -\infty}\n=\n_{\rm vac_A}, \qquad
    \lim_{\tilde{y}\to -\infty}\n=\n_{\rm vac_B},
    \label{eq:BC1}
\end{equation}
while $f_{\SW^-}$ and $f_{\LW^+}$ satisfy
\begin{equation}
    \lim_{\tilde{y}\to -\infty}\n=\n_{\rm vac_B}, \qquad
    \lim_{\tilde{y}\to -\infty}\n=\n_{\rm vac_A}.
    \label{eq:BC2}
\end{equation}
In Fig.~\ref{fig:wall_profile}, we represent the profile of the domain-wall solutions \eqref{eq:sol_SW} and \eqref{eq:sol_LW}.

The tension (energy per unit length) 
of the domain walls is given by
\begin{equation}
    T[f] = \int d\tily \left[
    \frac{1}{2}\left(\partial_{\tilde{y}} f\right)^{2}
    +s_2s_3\cdot|\kappa \cos \tilde{\vartheta}|  ~\partial_{\tilde{y}} f
    + s_1s_3\cdot |h| \sin f
    +\frac{\mu^{2}}{2} \sin ^{2} f 
    + \frac{h^2}{2\mu^2}\right ]
\end{equation}
where we have introduced the last constant term so that the tension of vacuum configurations $\sin f=-s_1s_3\gamma$ becomes zero.
Substituting the solutions into the tension, we obtain explicit expressions of the tensions 
of the short and long walls, respectively, as
\begin{align}
    &T_{\SW^\pm} \equiv T[f_{\SW^\pm}] = 2\mu \sqrt{1-\gamma^2}+4\left(-\mu\gamma\pm |\kappa\cos\tilth|\right)\arctan\sqrt{\frac{1-\gamma}{1+\gamma}} \,,
    \\
    &T_{\LW^\pm} \equiv T[f_{\LW^\pm}] = 2\mu \sqrt{1-\gamma^2}+4\left(\mu\gamma\pm |\kappa\cos\tilth|\right)\arctan\sqrt{\frac{1+\gamma}{1-\gamma}} \,.
\end{align}
This indicates that the domain walls $\SW^-$ and $\LW^-$ 
will spontaneously lie along the direction with $\tilde{\vartheta}=0$ 
unless it is pinned because the tension takes its lowest value at $\tilde{\vartheta}=0$ with given $\mu$ and $\kappa$. 
This is in contrast to the case with 
an out-of-plane easy-axis potential 
in which case the direction of 
a domain wall is arbitrary. 
In this paper, we consider an arbitrary value of $\tilde{\vartheta}$ for generality 
which would be achieved by 
pinning or applying suitable magnetic fields 
at the boundaries.
For a given value of $\tilth$, we have $T_{\SW^+}>T_{\SW^-}$ and $T_{\LW^+}>T_{\LW^-}$.
Moreover, one can easily find $T_{\LW^+} > T_{\SW^+}$.
Therefore, we have the following three possibilities in the order of domain-wall tension:
\begin{alignat}{3}
    &\text{Phase I}~:~
   T_{\LW^+} > T_{\LW^-}>T_{\SW^+}>T_{\SW^-}& \qquad \text{when} \qquad&  |\kappa\cos\tilth| < \mu\gamma \,.
    \label{eq:case3} 
    \\
    &\text{Phase II}~:~
    T_{\LW^+} > T_{\SW^+}\geq T_{\LW^-}>T_{\SW^-}
    & \qquad \text{when} \qquad
    &\mu\gamma\leq|\kappa\cos\tilth|<\frac{\pi\mu\gamma}{2\arctan\left(\dfrac{\gamma}{\sqrt{1-\gamma^2}}\right)}
    \,,
    \label{eq:case2}
    \\
    &\text{Phase III}~:~ 
    T_{\LW^+} > T_{\SW^+}>T_{\SW^-}\geq T_{\LW^-}
    & \qquad \text{when} \qquad
    & \frac{\pi\mu\gamma}{2\arctan\left(\dfrac{\gamma}{\sqrt{1-\gamma^2}}\right)} \leq |\kappa\cos\tilth|\,,
    \label{eq:case1}
\end{alignat}
\end{widetext}
For Phase II and III, $\SW^+$ is unstable because its tension is higher than that of $\LW^-$, which satisfies the same boundary condition with $\SW^+$. Similarly, $\LW^-$ is unstable in Phase I.

\begin{figure}[t]
    \centering
    \includegraphics[width=1\linewidth]{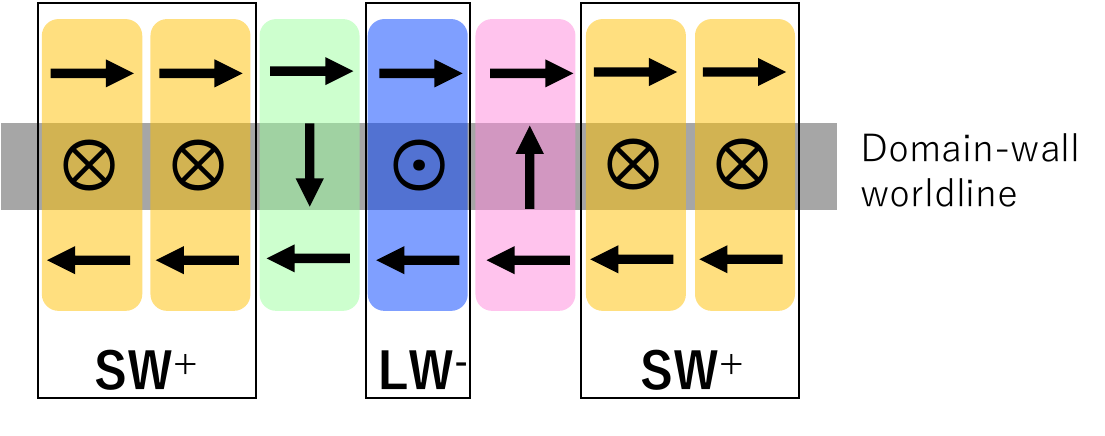}
    \caption{Schematic picture of a domain-wall skyrmion. 
    A domain wall lies horizontally.
    Specifically, in Phase I in Eq.~\eqref{eq:case3} where $E_{\SW^+}<E_{\LW^-}$, domain-wall skyrmions can be regarded as kinks of the magnetization vectors along the domain-wall worldline, separating two $\SW^+$ regions, and $\LW^-$ appears in the middle of the kink. }
    \label{fig:schematic}
\end{figure}

Domain-wall skyrmions are skyrmions confined to a domain wall and can be viewed as kinks on the domain wall. We show a schematic picture of a domain-wall skyrmion in Fig.~\ref{fig:schematic}.
In this case, the kink represents a twist in the phase $\phi$. 
The $\pi$ rotation of $\phi$ induces to the transformation $n_z\to -n_z$. It can approximately be regarded as 
a transition, e.g., 
from $\SW^+$ to $\LW^-$ along the domain-wall worldline. Therefore, roughly speaking, the kink connects $\SW^+ \to \LW^- \to \SW^+$ along the domain-wall worldline, which satisfy the same boundary condition \eqref{eq:BC1}. 
When the energy difference between $\SW^+$ and $\LW^-$ is small, as we shall see later, 
a lattice of 
domain-wall skyrmions is
the lowest energy single domain wall state satisfying the boundary condition \eqref{eq:BC1}.
In other words,  the CSL appears as the ground state inside the single domain wall.

In the same way, a domain-wall skyrmion can be constructed from the domain walls $\SW^-$ and $\LW^+$ satisfying the boundary condition \eqref{eq:BC2}.
The short domain wall $\SW^-$ has the minimum energy and the long domain wall $\LW^+$ has the largest energy in Phases I and II in Eqs.~\eqref{eq:case3} and \eqref{eq:case2}. 
Thus, one does not need to apply an external magnetic field 
to achieve this boundary condition, but 
the corresponding domain-wall skyrmion 
has relatively larger energy and cannot be the lowest energy state.

\section{Numerical analysis for domain-wall skyrmion crystals}
\label{sec:chain-numerics}

\begin{figure}
    \centering
    \includegraphics[width=\linewidth]
    {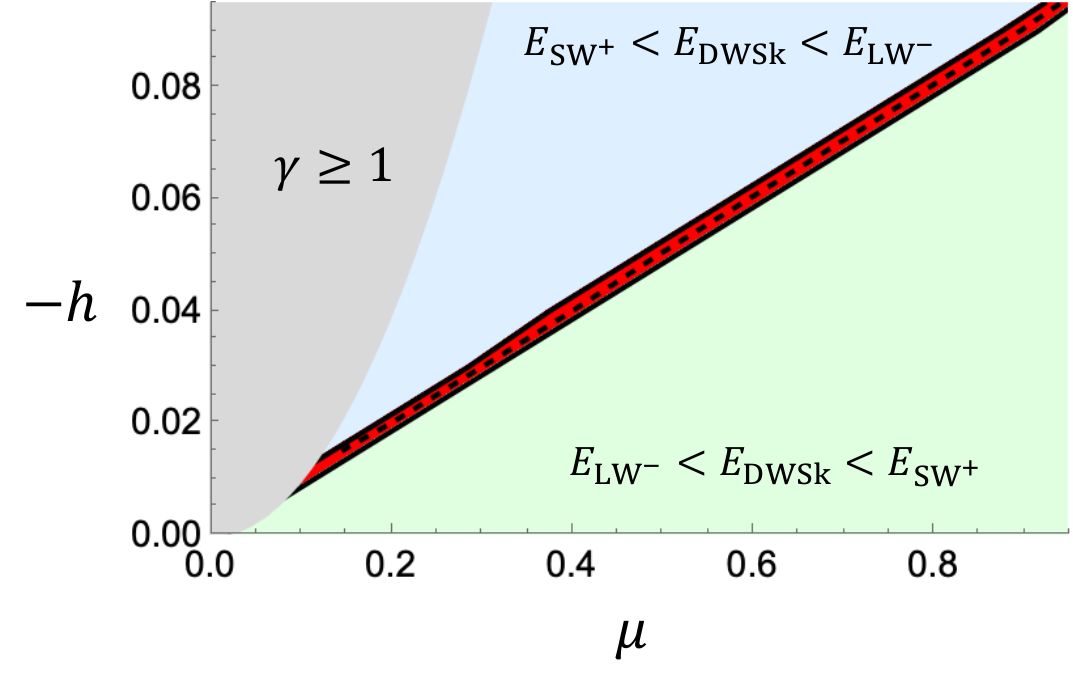}
    \caption{Phase diagram of the lowest energy single domain wall state (or the ground state inside a single domain wall). 
    The gray region ($\gamma \geq 1$) is excluded from the phase diagram since there is no domain wall in that region.
    Domain-wall skyrmion crystals appear as the lowest energy domain wall state in the red region between the two solid black lines. The lowest energy domain wall state in the blue region is $\SW^+$, and that in the green region is $\LW^-$.  
    On the dotted line, the energies of $\SW^+$ and $\LW^-$ coincide: $E_{\SW^+}=E_{\LW^-}$. In the region above the dotted line including the blue region, $E_{\SW^+}<E_{\LW^-}$, corresponding to Eq.~\eqref{eq:case3}. On the other hand, $E_{\SW^+}>E_{\LW^-}$ is held in the region below the dotted line including the green region. This region corresponds to Eqs.~\eqref{eq:case2} and \eqref{eq:case1}, where we do not distinguish the two phases because they have no difference as long as we focus on the relation between $E_{\SW^+}$ and $E_{\LW^-}$.
    Note that $E_{\SW^+}$ and $E_{\LW^-}$ are evaluated using numerical solutions representing $\SW^+$ and $\LW^-$ respectively, rather than the analytic solutions. For the numerical simulations, we used $\kappa = 0.1$. }
    
    \label{fig:phase_diagram_num}
\end{figure}

\begin{figure*}[!t]
    \centering
    \includegraphics[width=1.0\linewidth]{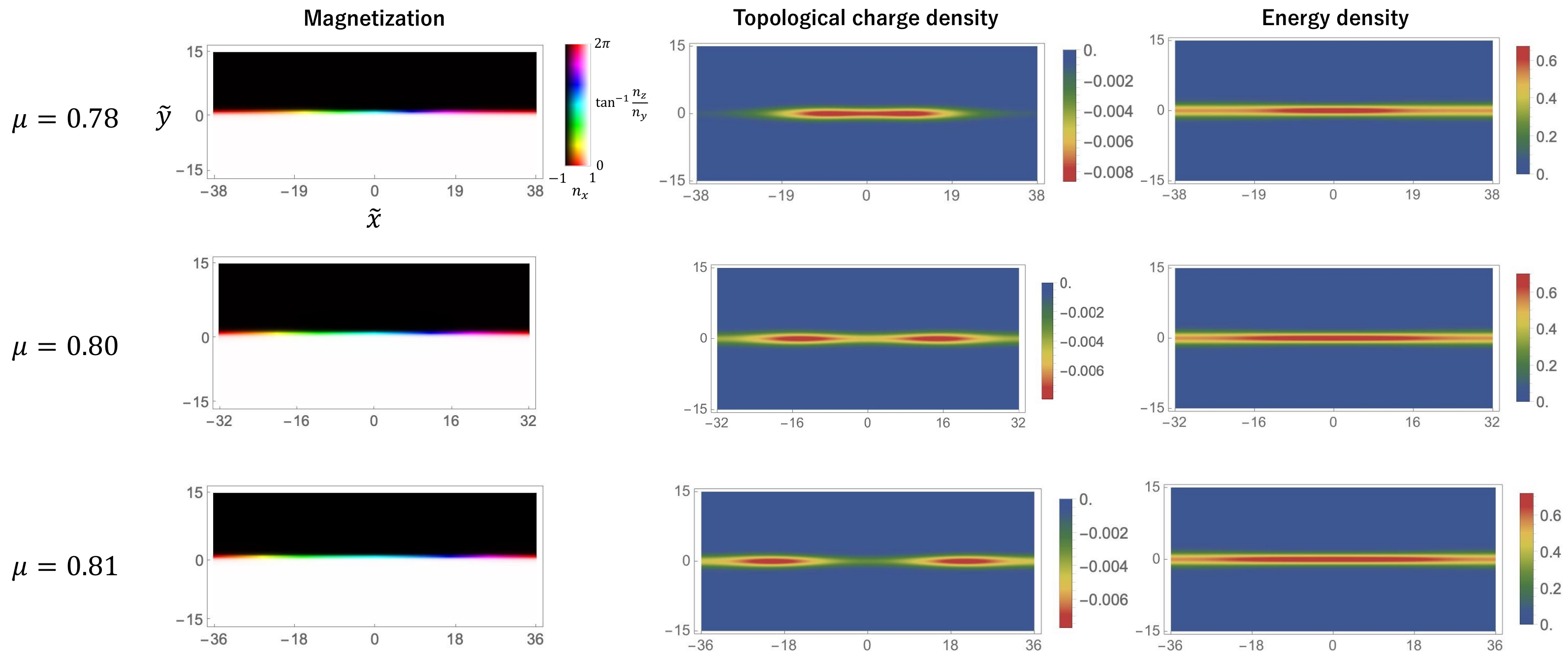}
    \caption{Domain-wall skyrmion crystals obtained by numerically solving the equation of motion.  These figures show the magnetization vector, topological charge density, and energy density of the configuration with a single period with $\mu=0.78,~0.80, \text{ and} ~0.81$.  The other parameters we used for the simulation are  $ \kappa= 0.1, h= -0.08,$ and $\tilde{\vartheta}=0$, and then the spin-spiral period $L_D=2\pi /|\kappa|=62.8$. 
    In the figures of the topological charge density, we observe that the lump representing a skyrmion at $\mu=1.0$ splits into two parts, each of which possesses approximately topological charge $-1/2$ and thereby can be regarded as a meron, when $\mu$ increases. On the other hand, the energy distribution shows only one peak independent of $\mu$, although the peak gradually delocalizes as $\mu$ increases. }
    \label{fig:chain_num}
\end{figure*}


In this section, we construct domain-wall skyrmion crystals 
 satisfying the boundary condition \eqref{eq:BC1} by numerically solving the equations of motion associated with the Hamiltonian \eqref{eq:bulk_theory}. 
 Here, domain-wall skyrmion crystals mean periodic arrays of skyrmions on a domain wall.
We find that domain-wall skyrmion crystals can have lower energy than the domain walls $\SW^+$ and $\LW^-$ in a certain parameter region. 
It implies that the domain-wall skyrmion crystals are the 
lowest energy state with the single domain wall structure
with the boundary condition \eqref{eq:BC1} that $\SW^+$ and $\LW^-$ satisfy. 
Such a 
boundary condition could be accomplished by applying a magnetic field on the edges of a magnetic material. 
For the numerical analysis, we restrict ourselves to the case $\tilde{\vartheta}=0$ 
where the energy of $\LW^-$ attains its lowest value and $\SW^+$ can exist as a saddle point
configuration.

The Euler-Lagrange equation to  solve is 
\begin{align}
        \partial_b^2 n_a 
    &+2\kappa\varepsilon_{abc}\partial_b n_c
    -h\delta_{az} +\mu^2\delta_{ax}n_x + \Lambda n_a = 0
        \label{eq:eom_skyrmion}
\end{align}
where $\Lambda$ is a Lagrange multiplier.
We solve Eq.~\eqref{eq:eom_skyrmion}
using a nonlinear conjugate gradient method with a finite difference approximation of fourth order. 
We impose 
the boundary condition \eqref{eq:BC1} 
as the Dirichlet boundary condition 
for the edge of the $\tilde{y}$ direction and the periodic boundary condition is used for the edge of the $\tilx$ direction.
The initial input is prepared as follows: for the profile $f$ in Eq.~\eqref{eq:n_parametrization}, we use smooth functions of $\tily$ corresponding to either $\SW^+$ or $\LW^-$, and for the phase $\phi$ we employ a smooth function of $\tilx$ monotonically varying $4\pi$. This implies that the system contains two periods of skyrmion crystal.
In the $\tilde{y}$ direction, we use 201 lattice points with lattice spacing $\Delta_{\tilde{y}}=0.05 \times L_D/2\pi$. 
For the $\tilde{x}$ direction, we employ 400 lattice points. The numerical simulations are performed with varying lattice spacing $\Delta_{\tilde{x}}$ to obtain configurations with proper periodicity.

In Fig.~\ref{fig:phase_diagram_num}, we show the phase diagram of the lowest energy single domain wall state. We find that domain-wall skyrmions appear as the lowest energy state in the red region.
Note that the domain-wall skyrmion crystal phase should ideally be larger than the red region in Fig.~\ref{fig:phase_diagram_num}, because the energy of the numerical solution is always over-estimated. We determine the phase boundary using domain-wall skyrmion crystals with a large but finite period, although the period of the skyrmion crystal diverges at the phase boundary.

Fig.~\ref{fig:chain_num} shows physical quantities of numerical solutions. 
At small $\mu$, the topological charge density exhibits a single peak in a single period. However, it gradually splits into two parts as $\mu$ increases. Each lump has the topological charge $Q=-1/2$ and thereby can be regarded as a meron. On the other hand, there exists only a single peak per one period in the energy density for any value of $\mu$. When we increase $\mu$, the peak in the energy density just delocalizes along the domain wall.
If one increases (decreases) $\kappa$, the period of the domain-wall skyrmion crystals becomes short (long), but the properties of the domain-wall skyrmion crystals do not change qualitatively.

Differences between our domain-wall skyrmion crystal and those obtained by a numerical simulation 
in Ref.~\onlinecite{Nagase:2020imn}  would be worth noting. 
In Fig.~\ref{fig:chain_num} b), we have two types of merons: the left one has $n_y<0$ and the right one has $n_y>0$.
Therefore, the domain-wall skyrmion crystal is a bipartite lattice of the merons $n_y<0$ and  $n_y>0$.
On the other hand, for domain-wall skyrmion (bimeron) crystals in Ref.~\onlinecite{Nagase:2020imn}, lumps of the topological charge density on a single domain wall are all equivalent and thereby the crystal is monopartite.

\section{Domain-wall Skyrmions in Moduli approximation }
\label{sec:moduli_approx}

In this section, we construct domain-wall skyrmions using the moduli approximation, also known as the Manton approximation~\cite{Manton:1981mp,Eto:2006pg,Eto:2006uw}. 
We find that the approximation qualitatively reproduces the features of numerical solutions.

\subsection{Single domain-wall skyrmion}
\label{subsec:single}

In this subsection, we analytically construct single domain-wall skyrmions using the moduli approximation.  
By evaluating their energy, we discuss the lowest energy structure on the domain wall.

Moduli approximation is to consider low-energy fluctuation around a soliton while neglecting the back-reactions of the fluctuation.
To describe the low-energy fluctuation, moduli parameters of solitons are promoted to fields depending on coordinates along the solitons, and an effective theory for the fields is constructed by integrating the Lagrangian or Hamiltonian over co-dimension of the solitons.
In our case, the low-energy fluctuations of domain walls are associated with the position moduli $\chi$ and the phase $\phi$.
Note that since the phase $\phi$ contributes to the domain wall's energy, it is not precisely a moduli parameter. It is, however, regarded as a quasi-moduli parameter because Eqs.~\eqref{eq:sol_SW} and \eqref{eq:sol_LW} satisfy the equation of motion for any value of $\phi$, implying that the existence of the domain wall is independent of $\phi$.

To apply the moduli approximation, we promote the (quasi-)moduli parameters $\chi$ and $\phi$ to fields depending on the coordinates along the wall:
\begin{align}
& \chi=\chi(\tilde{x}) , \\
& \phi=\varphi(\tilde{x}) 
+ s_3\frac{\pi}{2}.
\label{eq:promotion_phi}
\end{align}
However, Eq.~\eqref{eq:promotion_phi} causes the problem that the asymptotic value of $\n$ at $\tilde{y}=\pm \infty$ for the short and long walls changes. In order to resolve the problem, instead of Eqs.~\eqref{eq:sol_SW} and \eqref{eq:sol_LW}, we approximately use the profile function 
in the limit $h \to 0$:
\begin{align}
    \lim_{h\to0} f_{\SW^\pm} 
    =&\pm 2\arctan\left[
    \exp\left(s_2s_3 \{\mu\tily-\chi\}\right)
    \right]
    \notag\\
    &~
    -(\pm 1 + s_1s_3)\frac{\pi}{2} \, ,
    \label{eq:SW_h0}
\end{align}
\begin{align}
    \lim_{h\to0} f_{\LW^\pm}    
    =&\pm 2\arctan\left[
    \exp\left(s_2s_3  \{\mu\tily-\chi\}\right)
    \right]
    \notag\\
    &~
    -( \pm 1 - s_1s_3)\frac{\pi}{2}
    \label{eq:LW_h0}
\end{align}
The tension of the configurations \eqref{eq:SW_h0} and \eqref{eq:LW_h0} with $\chi=\varphi=0$ are given respectively as
\begin{align}
    \lim_{h\to 0} T[f_{\SW^\pm}] = \lim_{h\to 0} T[f_{\LW^\pm}]  =  2|\mu| \pm \pi|\kappa \cos\tilde{\vartheta}| \ .
\end{align}
In this paper, specifically, we employ $\LW^-$ in the limit $h\to0$ as the approximated seed domain-wall configuration in order to compare the results of the moduli approximation and numerical results.
Therefore, we write
\begin{align}
    &f_\DW = \lim_{h\to0} f_{\LW^-} 
    \notag\\
    &=-2\arctan\left[
    \exp\left(s_2s_3 \{\mu \tily-\chi\}\right)
    \right]
     +( 1 + s_1s_3)\frac{\pi}{2}
     \label{eq:DW}
\end{align}
The configurations $\DW$ describe a domain wall when $s_2s_3=-1$ and an anti-domain wall when $s_2s_3=+1$.
In other words, the function $f_{\DW}$  monotonically increases (decreases) in $\tilde{y}$ for $s_2s_3=-1 (+1)$. 
The tension of $\DW$ with $\chi=\varphi=0$ is given by
\begin{align}
    T_{\DW} &\equiv T[f_{\DW}] =  2|\mu| - \pi|\kappa \cos\tilde{\vartheta}| \ .
    \label{eq:dw_tension}
\end{align}
Clearly, it takes the lowest value when we choose the minus sign with $\tilth=0$.

The effective energy for the fluctuation is given by the integral of the Hamiltonian density over the coordinates normal to the domain wall and can be cast into the form
\begin{align}
{\cal E}_\text{eff} \equiv&  \int d \tilde{y}~ \calH 
\notag\\
=&\frac{2}{|\mu|}\left[\frac{1}{2}\left(\partial_{\tilde{x}} \chi - A\cos\varphi \right)^2
\right. \notag\\
& \left. \qquad \quad
+ \frac{1}{2}\left(\partial_{\tilde{x}} \varphi\right)^{2}
-\kappa \cos \tilde{\vartheta} \partial_{\tilde{x}} \varphi
+V(\varphi)
 \right]
 \label{eq:edens_eff}
\end{align}
where $V(\varphi)$ is the potential given by
\begin{align}
V(\varphi)=|B|-B \cos \varphi+\frac{A^{2}}{2}\left(1-\cos ^{2} \varphi\right) \ .
\end{align}
Here, the parameters $A$ and $B$ are given by
\begin{align}
\begin{aligned}
& A \equiv \frac{\pi \kappa}{2} s_2 \sin \tilde{\vartheta}, 
\\
& B \equiv \frac{\pi}{2} (|\kappa \mu \cos \tilde{\vartheta}|- |h|),
\label{eq:AB}
\end{aligned}
\end{align}
where we have omitted constant terms depending on $\tilde{\vartheta}$. 
The equations of motion associated with the effective energy \eqref{eq:edens_eff} are given by
\begin{align}
& \partial_{\tilde{x}}^{2} \varphi=B \sin \varphi+\frac{A^{2}}{2} \sin 2 \varphi 
\label{eq:eom_phi}\\
& \partial_{\tilde{x}} \chi=  A \cos \varphi.
\label{eq:eom_chi}
\end{align}
Eq.~\eqref{eq:eom_chi} is of the Bogomol'nyi type. One can obtain it by integrating the Euler-Lagrange equation with respect to $\chi$ once.
Eq.~\eqref{eq:eom_phi} is the double sine-Gordon equation that can be obtained from the Euler-Lagrange equation with respect to $\varphi$ by eliminating its $\chi$-dependence using Eq.~\eqref{eq:eom_chi}.
Note that 
the field $\chi$ decouples with $\varphi$ when $\tilde{\vartheta}=0$ because $A \propto \sin\tilde{\vartheta}=0$.

The vacua of the effective energy is given by the constant field
\begin{align}
\varphi=\left\{\begin{array}{cl}
2 l \pi & \mbox{ for } B>0 \\
l \pi & \mbox{ for } B=0 \\
(2 l+1) \pi & \mbox{ for } B<0 .
\end{array}\right.
\end{align}
with $l \in {\mathbb Z}$.
When $A\neq0$, the field $\chi$ for the vacuum is linear in $\tilde{x}$, i.e.,
\begin{equation}
    \chi = \sign(B) A  \tilde{x} + \const. \ .
\end{equation}
This implies that the domain wall is straight and  slanted.

Now, let us discuss kink solutions in the domain-wall effective theory.
A single kink connects two neighboring vacua.
Integrating Eq.~\eqref{eq:eom_phi} once, we obtain 
\begin{align}
\begin{aligned}
\partial_{\tilde{x}} \varphi= s_2 \sqrt{2(V(\varphi)+\Delta)}
\label{eq:pd_phi}
\end{aligned}
\end{align}
with an integration constant $\Delta$.
The sign is chosen as the first derivative term gives a negative contribution. 
When $\Delta=0$, it is nothing but the Bogomol'nyi equation for the double sine-Gordon model that a single kink solution satisfies.

Kink solutions in this effective theory describe single domain-wall skyrmions in the bulk theory~\eqref{eq:bulk_theory}.
When $B\neq0$, we impose the boundary condition
\begin{align}
\lim _{\tilde{x} \rightarrow-\infty} \partial_{\tilde{x}} \varphi=0,
\qquad 
\lim _{\tilde{x} \rightarrow-\infty} \cos \varphi=\sign(B)
\end{align}
Then, Eq.~\eqref{eq:pd_phi} can be explicitly rewritten as
\begin{align}
& \partial_{\tilde{x}} \varphi= s_2\sqrt{2 |B|-2B\cos \varphi+A^{2}\sin ^{2} \varphi} .
\end{align}
A single kink solution is given by
\begin{widetext}
\begin{align}
\varphi=-2s_2 \cdot \arccos \left[ \sqrt{\frac{|B| }{A^{2}+|B| \cosh ^{2}\left(\sqrt{A^{2}+|B|} \tilde{x}+\alpha\right)}}\sinh \left(\sqrt{A^{2}+|B|} \tilde{x}+\alpha\right)\right] 
+(4l+1-\sign(B))\frac{\pi}{2}
\label{eq:phi_kinksol_case1}
\end{align}
with a constant $\alpha$.  Using the solution \eqref{eq:phi_kinksol_case1}, Eq.~\eqref{eq:eom_phi} can be 
rewritten as
\begin{align}
\begin{aligned}
\partial_{\tilde{x}} \chi 
 =-\sign(B) ~ A\left(1 - \frac{2|B| \sinh ^{2}\left(\sqrt{A^{2}+|B|} \tilde{x}+\alpha\right)}{A^{2}+|B| \cosh ^{2}\left(\sqrt{A^{2}+|B|} \tilde{x}+\alpha\right)}\right) .
\end{aligned}
\end{align}
Integrating both sides of this equation, one finally obtains 
the single kink solution 
\begin{align}
\begin{aligned}
\chi=\sign(B) \left(  \frac{A}{\sqrt{A^{2}+|B|}}\left(\sqrt{A^{2}+|B|} \tilde{x}+\alpha\right) 
 -2 \operatorname{arctanh}\left[\frac{A}{\sqrt{A^{2}+|B|}} \tanh \left(\sqrt{A^{2}+|B|} \tilde{x}+\alpha\right)\right] \right) +\beta
\end{aligned}
\end{align}
with a constant $\beta$.
The energy of this kink solution is obtained as
\begin{align}
E_{\text{eff}}=\int d \tilde{x} ~{\cal E}_{\text {eff }}
=\frac{8}{|\mu|}\left\{\sqrt{A^{2}+|B|}+\frac{|B|}{A} \operatorname{arctanh}\left(\frac{A}{\sqrt{A^{2}+|B|}}\right) - \frac{\pi }{2} |\kappa\cos \tilde{\vartheta}| \right\}.
\label{eq:kinkenergy_case1}
\end{align}
\end{widetext}
When $A=0=\tilde{\vartheta}$, the energy can simply be written as 
\begin{equation}
    E_\text{eff}=\frac{8}{|\mu|}\left(\sqrt{2\pi||\kappa\mu|-|h||}-\frac{\pi}{2}|\kappa|\right) \ .
\end{equation}

On the other hand, 
in the case with $B=0$, we impose boundary conditions of the form
\begin{align}
    \lim _{\tilde{x} \rightarrow-\infty} \partial_{\tilde{x}} \varphi=0,\qquad 
    \lim _{\tilde{x} \rightarrow-\infty} \cos \varphi= \pm 1
\end{align}
Then, Eq.~\eqref{eq:pd_phi} reads
\begin{align}
\partial_{\tilde{x}} \varphi= s_2 ~ |A\sin\varphi| .
\end{align}
This is nothing but the Bogomol'nyi equation for the sine-Gordon equation. The single kink solution is given by
\begin{align}
\begin{aligned}
\varphi=2 s_2\cdot  \arctan [\exp ( |A| \tilde{x}+\alpha)] + l\pi .
\label{eq:phi_kinksol_case2}
\end{aligned}
\end{align}
Similarly to the case with $B\neq0$, we substitute Eq.~\eqref{eq:phi_kinksol_case2} into Eq.~\eqref{eq:eom_chi} 
to obtain
\begin{align}
\begin{aligned}
\partial_{\tilde{x}} \chi & = (-1)^{l+1} A \tanh ( |A| \tilde{x}+\alpha) \ .
\end{aligned}
\end{align}
The integration over both sides gives us the solution
\begin{align}
\chi= (-1)^{l+1} \sign(A) \log [\cosh ( |A| \tilde{x}+\alpha)]+\beta  \ .
\end{align}
The energy difference between this solution and the vacuum is
\begin{align}
\begin{aligned}
E_\text{eff}  =\frac{4}{|\mu|}
\left( |A| - \frac{\pi }{2} |\kappa\cos \tilde{\vartheta}| \right) \ .
\label{eq:kinkenergy_case2}
\end{aligned}
\end{align}
This is a half of Eq.~\eqref{eq:kinkenergy_case1} with $B=0$.

The kink energies in Eqs.~\eqref{eq:kinkenergy_case1} and \eqref{eq:kinkenergy_case2} are not positive definite due to their last term coming from the first derivative topological term in Eq.~\eqref{eq:edens_eff}. When the kink energy is positive, the ground state on the domain-wall theory is 
ferromagnetic, i.e., uniform. 
On the other hand, 
when the kink energy is negative, 
the ground state is no longer uniform and 
it is a CSL state, 
corresponding to a one-dimensional skyrmion crystal 
in the bulk.
We will study this case in the following subsection.

\begin{figure}
    \centering
    \includegraphics[width=1.0\linewidth]{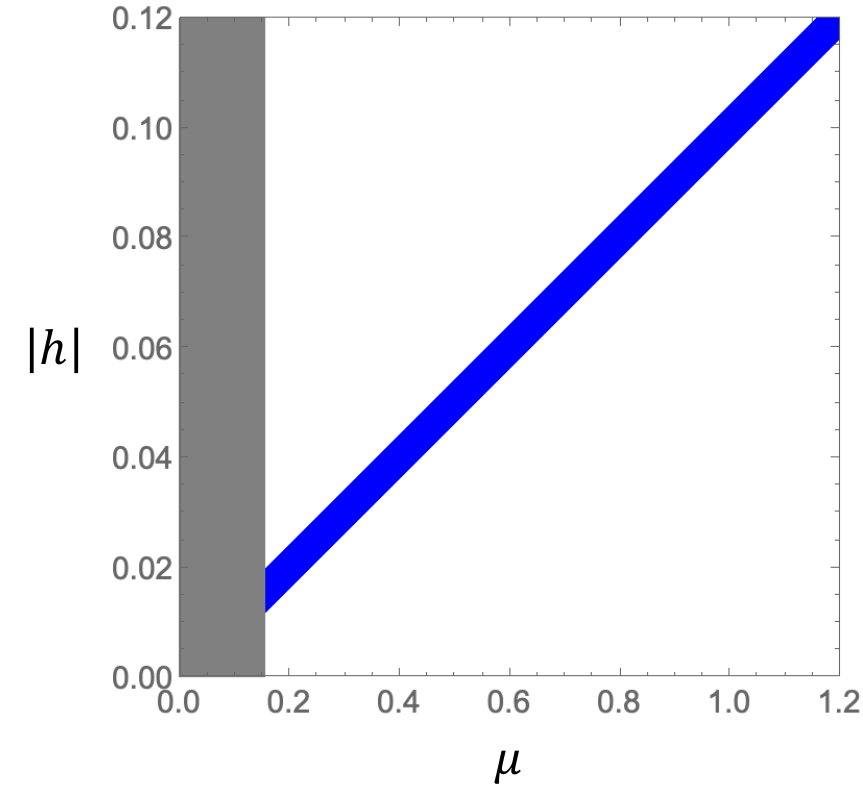}
    \caption{Phase diagram for the effective theory \eqref{eq:edens_eff}. The blue region represents the CSL phase on the domain wall given by $8||\kappa\mu| - |h|| < \pi \kappa^2$, where domain-wall skyrmion crystals are preferred over the standard domain wall~\eqref{eq:DW} in the bulk theory. We exclude the gray region corresponding to $4\mu^2 < \kappa^2\pi^2$, where the ground state in the bulk theory is no longer ferromagnetic with negative domain-wall tension~\eqref{eq:dw_tension}, because we are restricting ourselves to the study of the ferromagnetic phase as we wrote above. For this figure, we used $\kappa = 0.1$. }
    \label{fig:phase_diagram_analytical}
\end{figure}

We show the phase diagram on the domain wall 
in Fig.~\ref{fig:phase_diagram_analytical}. 
Note that in the case of an out-of-plane easy-axis anisotropy instead of the in-plane anisotropy, the energy of kinks in a domain-wall effective theory is always positive, and the system does not possess a CSL phase on the domain wall.

\subsection{Skyrmion crystal phase on a domain wall}
\label{subsec:chain}

\begin{figure*}[!t]
    \centering
    \includegraphics[width=1.0\linewidth]{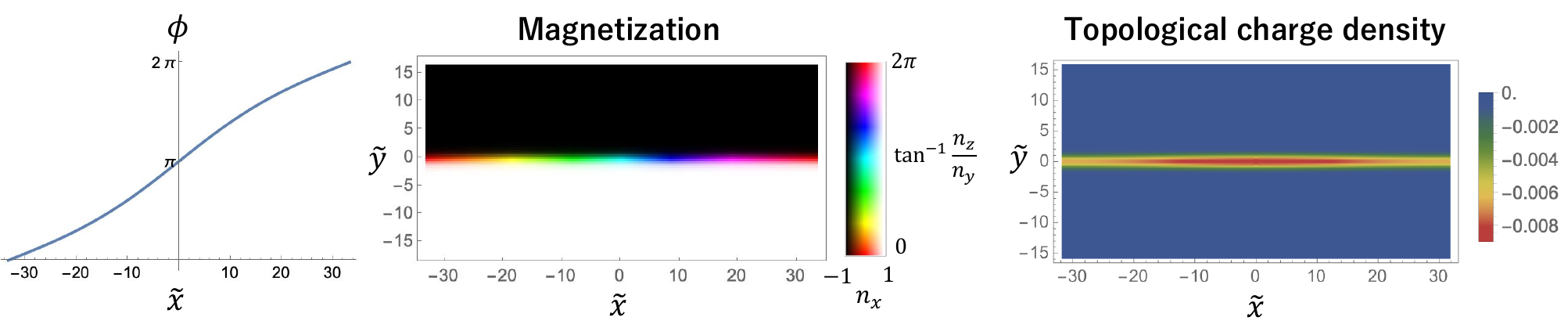}
    \caption{Domain-wall skyrmion crystal in the effective theory with $A=0$ 
    and $B\neq 0$ (analytical solution). 
    From left to right, the figures represent the phase, profile of the magnetization vector, and topological charge density, respectively.
For these figures, we use the analytic solution Eq.~\eqref{eq:phi_kinksol_case1} with $B=0.0005\times\pi$ corresponding to the parameter set $(\kappa,\mu, h) = (0.1, 1.01, -0.1)$. }
    \label{fig:chain_effth_A=0}
\end{figure*}

\begin{figure*}[!t]
    \centering
    \includegraphics[width=1.0\linewidth]{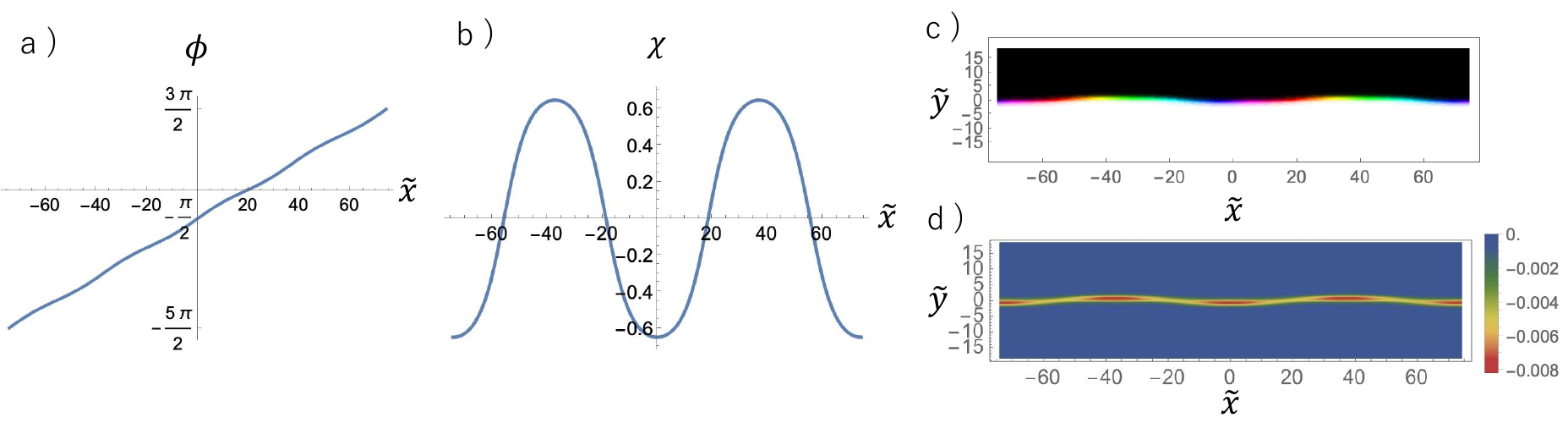}
    \caption{Domain-wall skyrmion crystal in the effective theory with $B=0$ (analytical solution): a) phase $\phi$, b) profile $\chi$, c) magnetization vector, and d) topological charge density. The color scheme for the magnetization vector is the same as in Fig.~\ref{fig:chain_effth_A=0}. For the solutions, we used $A=0.025\times\pi$ corresponding to the parameter set $(\kappa, \tilde{\vartheta}, \mu) = (0.1, \pi/6, 1.01)$}
    \label{fig:chain_effth_B=0}
\end{figure*}

\begin{figure*}[!t]
    \centering
    \includegraphics[width=1.0\linewidth]{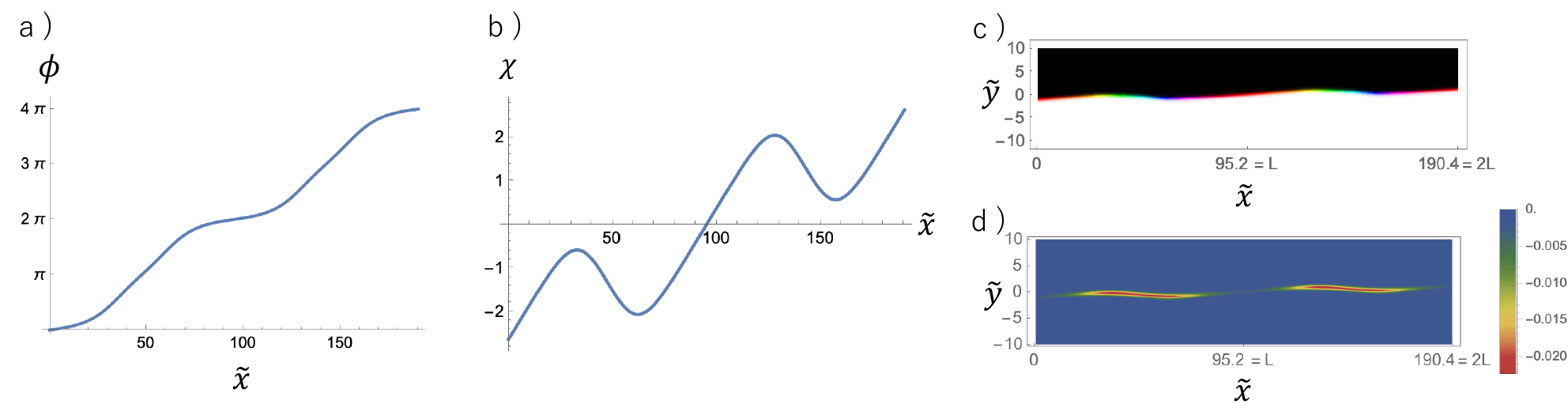}
    \caption{Domain-wall skyrmion crystal in the effective theory with $A, B\neq 0$ (numerical solution): a) phase $\phi$, b) profile $\chi$, c) magnetization vector, and d) topological charge density. The color scheme for the magnetization vector is the same as in Fig.~\ref{fig:chain_effth_A=0}. For the solution, we used the parameter set $(\kappa, \tilde{\vartheta}, \mu, h)=(0.1, \pi/6, 1.0, -0.085)$. }
    \label{fig:chain_effth_nonzero}
\end{figure*}

The domain-wall skyrmion crystals can be described by CSL solutions in the effective theory on the domain wall.
The CSL with an appropriate period is the ground state in the effective theory.
The period $L$ of the CSL is defined from Eq.~\eqref{eq:pd_phi} as
\begin{align}
L=\int_{0}^{2 \pi} \frac{d \varphi}{\sqrt{2(V(\varphi)+\Delta)}} \ .
\end{align}
We specify the parameter $\Delta$ providing the appropriate period.
We can write the energy per unit length as
\begin{align}
\begin{aligned}
& \bar{E}_\text{eff} \equiv E_\text{eff}/L \\
& 
=\frac{2}{|\mu|}\frac{1}{L}\left[ \int_{0}^{2 \pi} d \varphi \sqrt{2(V(\varphi)+\Delta)} \right.
\\
&\qquad\qquad\qquad\qquad\qquad\left.
-L \Delta-2 \pi|\kappa \cos \tilde{\vartheta}|\right] \ ,
\end{aligned}
\end{align}
where we utilized Eq.~\eqref{eq:pd_phi}.
It follows that
\begin{align}
\begin{aligned}
\frac{\partial \bar{E}_\text{eff}}{\partial \Delta}
= & \frac{2}{|\mu|} \frac{1}{L^{2}} \frac{\partial L}{\partial \Delta}
\\
&\times\left[2 \pi|\kappa \cos \tilde{\vartheta}|-\int_{0}^{2 \pi} d \varphi \sqrt{2(V(\varphi)+\Delta)}\right].
\end{aligned}
\end{align}
Since $L$ is a monotonic function in $\Delta$, we have $\partial L/\partial \Delta \neq 0$.
Therefore, the least-energy condition for $\Delta$ is given by
\begin{align}
\int_{0}^{2 \pi} d \varphi \sqrt{2(V(\varphi)+\Delta)}=2 \pi|\kappa \cos \tilde{\vartheta}| \ .
\label{eq:condition_delta}
\end{align}
The ground state in the effective theory is given by the CSL solution with $\Delta$ satisfying Eq.~\eqref{eq:condition_delta}, and its energy is $\bar{E}_\text{eff}=-2\Delta/ (L|\mu|)$. 

We can analytically derive CSL solutions if $A=0$ or $B=0$ in which case Eq.~\eqref{eq:eom_phi} reduces to the sine-Gordon equation. Otherwise, we have to rely on a numerical method.
\\
\\
\noindent
 \textbf{i) The case of $\bm{A=0, ~B \neq 0}$.} 
 
 From Eq.~\eqref{eq:AB}, one can see that 
 $A=0$ is achieved by $\tilde \theta =0$ mod $\pi$, implying that 
the domain wall lies in the most energetically preferable direction\footnote{The effective theory is invariant under the simultaneous parameter transformation $\tilde{\vartheta}=0\to \pi, h\to -h$. For simplicity, we only consider the case $\tilde{\vartheta}=0$ and allow $h$ to take either positive or negative values. }. 
 In such a case, the equation for $\varphi$ reduces to the sine-Gordon equation with the $2\pi$ period for $\varphi$ and Eq.~\eqref{eq:pd_phi} can be cast into the form
\begin{align}
\begin{aligned}
& \partial_{\tilde{x}} \varphi= s_2 \sqrt{2(|B|-B \cos \varphi+\Delta)} \ .
\end{aligned}
\end{align}
The solution is given in terms of the Jacobi amplitude function as
\begin{align}
\begin{aligned}
    \varphi=2 &s_2 \operatorname{am}\left(  \frac{\sqrt{|B|}}{\lambda} \tilde{x}+\alpha, \lambda\right)
    +(1+\operatorname{sign}(B)) \frac{\pi}{2}
    \label{eq:phi_CSLsol_case1}
\end{aligned}
\end{align}
with the elliptic modulus $\lambda=\sqrt{\frac{2|B|}{2|B|+\Delta}}$ and a constant $\alpha$.
One can deduce the least-energy condition for the modulus $\lambda$ from Eq.~\eqref{eq:condition_delta} as
\begin{align}
\begin{aligned}
\frac{\E(\lambda)}{\lambda}=\frac{\pi}{4}\frac{|\kappa\cos \tilde{\vartheta}|}{\sqrt{|B|}}
\end{aligned}
\end{align}
with the elliptic integral of the second kind $\E(\lambda)$. Here, we have used 
\begin{align}
\begin{aligned}
 \int_{0}^{2 \pi} d \varphi \sqrt{|B|-B \cos \varphi+\Delta} =4 \sqrt{2|B|+ \Delta} ~ \E(\lambda) .
\end{aligned}
\end{align}
In this case, $\chi$ does not depend on the coordinate, 
implying that the domain wall is straight.
In Fig.~\ref{fig:chain_effth_A=0} we plot quantities of the solution \eqref{eq:phi_CSLsol_case1} 
: phase $\phi$, magnetization vector $\n$, and topological charge density.
The integral of the topological charge density over a single period gives $-1$ for the solutions.
\\
\\
\noindent
\textbf{ii) The case of $\bm{A \neq 0, B=0}$.}

From Eq.~\eqref{eq:AB},
one can observe that the condition $B=0$ can be achieved 
by fine-tuning the Zeeman magnetic field 
as
\begin{equation}
 |h| = - |\kappa \mu \cos \tilde{\vartheta}|.
\end{equation}
From Eq.~\eqref{eq:pd_phi}, we obtain
\begin{align}
\partial_{\tilde{x}} \varphi= s_2 \sqrt{A^{2}\left(1-\cos ^{2} \varphi\right)+2\Delta} .
\end{align}
Integrating this equation gives 
\begin{align}
\begin{aligned}
\varphi=s_2\operatorname{am}\left(  \frac{|A|}{\lambda} \tilde{x}+\alpha, \lambda\right)-\frac{\pi}{2}
\label{eq:phi_CSLsol_case2}
\end{aligned}
\end{align}
with the elliptic modulus $\lambda=\sqrt{\frac{A^{2}}{A^{2}+2 \Delta}}$.
Using Eq.~\eqref{eq:phi_CSLsol_case2}, we get the solution of Eq.~\eqref{eq:eom_chi} of the form
\begin{align}
\begin{aligned}
    \chi= - &s_2\sign(A) \\
    &\times \arctan \left[\lambda \operatorname{cd}\left(\frac{|A|}{\lambda}\tilx+\alpha, \lambda\right)\right]+\beta.
\end{aligned}
\label{eq:chi_CSLsol_case2}
\end{align}
In this case, we have 
\begin{align}
\begin{aligned}
& \int_{0}^{2 \pi} d \varphi \sqrt{A^{2}\left(1-\cos ^{2} \varphi\right)+2 \Delta}=\frac{4|A|}{\lambda} \E(\lambda).
\end{aligned}
\end{align}
Therefore Eq.~\eqref{eq:condition_delta} indicates
\begin{align}
\frac{\E(\lambda)}{\lambda}=\frac{\pi}{2}\frac{|\kappa|}{|A|}|\cos\tilde{\vartheta}| = |\cot\tilde{\vartheta}|.
\end{align}

In Fig.~\ref{fig:chain_effth_B=0} we plot the solution \eqref{eq:phi_CSLsol_case2} with \eqref{eq:chi_CSLsol_case2}.
One can observe that the domain wall is periodically oscillating with a half period of the case i), 
and each skyrmion is split into two merons.
\\
\\
\noindent
\textbf{iii) The case of $\bm{A \neq 0, B \neq 0}$.} 

In this case, we cannot analytically obtain CSL solutions. We construct kink lattices using the following numerical procedure:
\begin{itemize}
    \item[1.] 
    For given parameter set $(\kappa, \tilde{\vartheta}, \mu, h)$, determine $\Delta$ via Eq.~\eqref{eq:condition_delta} with the Newton method.

    \item[2.] 
    With $\Delta$ satisfying Eq.~\eqref{eq:condition_delta}, solve Eq.~\eqref{eq:eom_phi} with the initial condition $\varphi(0)=\varphi_{\text {vac }}$ by the Runge-Kutta method while $\varphi$ changes $2 \pi$.

    \item[3.]
    Solve Eq.~\eqref{eq:eom_chi} with $\chi{(0)}=0$ by the Runge-Kutta method. 
\end{itemize}

In Fig.~\ref{fig:chain_effth_nonzero}, we plot the numerical solution. 
In this case, the domain wall fluctuates like that in the case of 
$A \neq 0$ and $B = 0$.
With the oscillation, 
the mean position $\chi$ over a period is linearly increasing in $\tilde x$ 
implying that the domain wall is slanted 
[see Fig.~\ref{fig:chain_effth_nonzero} b)]. 
This is because a line of a constant $\tilde y$ ($\chi =0$) is 
not energetically the most preferable direction of 
the domain wall. 
One also can see that 
each skyrmion is spread and is about to split into two merons.

\subsection{Comparison of results by the full numerical simulations and moduli approximation}
\label{subsec:comparison}

\begin{figure}[t]
    \centering
    \includegraphics[width=1\linewidth]{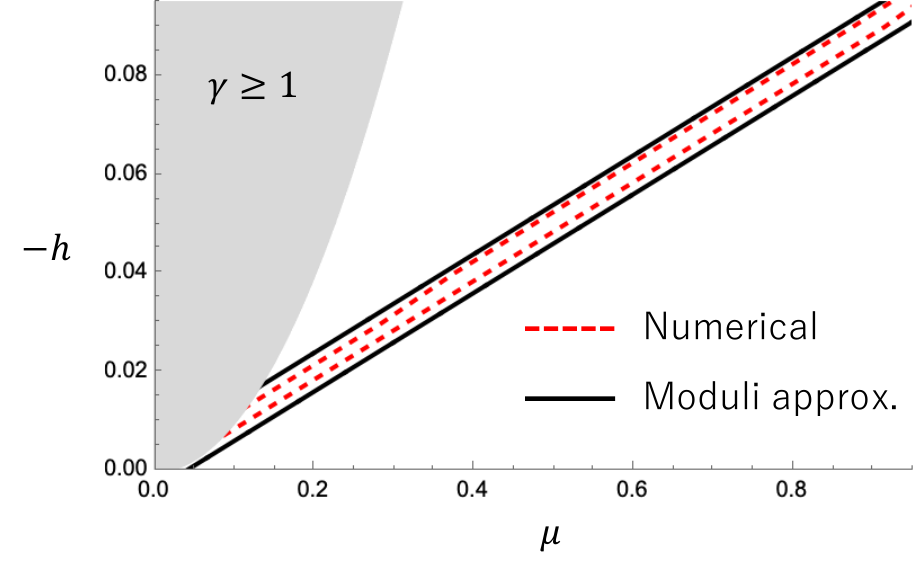}
    \caption{Comparison of the parameter regions in which domain-wall skyrmion crystals appear as the ground state on a single domain wall obtained by the numerical simulations and moduli approximation.
    The region between the two red dotted lines is the one obtained by the full numerical solution shown in Fig.~\ref{fig:phase_diagram_num}.
    On the other hand, the black solid lines show the boundary between the CSL and uniform phases in the effective theory. Inside the two black lines is the region where domain-wall skyrmion crystals appear as the ground state on a single domain wall obtained by the moduli approximation. For the plot, we used $\kappa=0.1$ and $\tilde{\vartheta}=0$, which are also used in Figs.~\ref{fig:phase_diagram_num} and \ref{fig:phase_diagram_analytical}}
    \label{fig:comparison}
\end{figure}

We compare the results of the full numerical simulation given in Sec.~\ref{sec:chain-numerics} and those of the moduli approximation discussed in this section.
In Fig.~\ref{fig:comparison}, we show the parameter regions where domain-wall skyrmion crystals appear as the lowest energy single domain wall state obtained by the two methods.
We find that they have a quantitatively good agreement.
The region obtained by the numerical simulation must be smaller than the actual one. The energy of the numerical solution is always over-estimated because we determine the phase boundary using a skyrmion crystal with a large but finite period, although the period of the skyrmion crystal diverges at the phase boundary. Therefore, we can conclude that the moduli approximation qualitatively reproduces the phase diagram on a domain wall quite well.

Let us compare the period length of the domain-wall skyrmion crystals. That obtained by the numerical simulation becomes slightly longer than that of the analytical solution in the moduli approximation. Specifically, for the parameter set $(\kappa,\mu,h)=( 0.1,0.78,-0.08)$, the period length of the numerical solution is approximately 1.21$L_D$, while that of the analytical solution is around 1.06$L_D$. 

The most remarkable difference between the results of the two methods is the decomposition of a domain-wall skyrmion into two merons.
In the moduli approximation, we can always observe only a single peak per one period representing a $Q=-1$ skyrmion at $\tilde{\vartheta}=0$.
The split of the domain-wall skyrmion occurs only when $\tilde{\vartheta}\neq 0$.
On the other hand, as shown in Fig.~\ref{fig:chain_num}, for the full numerical simulation, a domain-wall skyrmion is split into two merons even at $\tilde{\vartheta}=0$.

These results indicate that although the moduli approximation fails to capture the split of the topological charge of the domain-wall skyrmion into merons, it successfully reproduces significant properties of domain-wall skyrmion crystals, such as their emergence as the lowest energy state with a single domain wall structure, the absence of domain wall bending observed in chiral magnets with an out-of-plane easy-axis anisotropy \cite{Amari:2023gqv,Amari:2023bmx}, etc. 
The reason why the moduli approximation fails to describe some properties of the domain-wall skyrmion crystals will be that the effective theory \eqref{eq:edens_eff} is constructed based on the domain-wall solution in the absence of the Zeeman interaction,
unlike the case of out-of-plane easy-axis anisotropy
in Ref.~\cite{Amari:2023bmx} in which 
the Zeeman interaction is absent 
both in the effective theory and numerical solutions.
The Zeeman interaction makes the difference between the short and long walls. 
We cannot distinguish them within the moduli approximation. That is probably the main reason of mismatch between the two methods.

\section{Summary and Discussion}
\label{sec:discussion}

We have investigated domain-wall skyrmions 
in chiral magnets with 
an in-plane
easy-axis anisotropy 
as well as
an out-of-plane Zeeman magnetic field, 
instead of the out-of-plane easy-axis 
anisotropy studied in 
the previous works.
The direction of a single straight domain wall has an energetically preferable direction. 
First, we have performed full numerical simulations
of the configuration with the boundary condition \eqref{eq:BC1}
and 
obtained the phase diagram of the lowest energy single domain wall state in Fig.~\ref{fig:phase_diagram_num}.
Even though the bulk (without any specific boundary condition) 
is in the ferromagnetic phase, the ground state 
on a single domain wall state
is either the uniform ferromagnetic phase 
or the CSL phase on the domain wall.
The latter corresponds to 
a skyrmion crystal along the domain line. 
Second, 
we have studied this system analytically 
by using the moduli approximation. 
We have constructed 
the effective theory on a single domain wall 
in the ferromagnetic phase 
 with perturbatively treating the Zeeman magnetic field,  
and obtained a chiral double sine-Gordon model. 
By constructing double sine-Gordon solitons,
we have obtained the phase diagram 
of the domain-wall effective theory
in Fig.~\ref{fig:phase_diagram_analytical}, 
which we found is in a good agreement 
with 
that of full numerics, 
as shown in Fig.~\ref{fig:comparison}.
While 
 the moduli approximation captures 
the qualitative behaviors of numerical solutions, 
the latter exhibits more precise configurations 
of domain-wall skyrmions 
such as a decomposition of the topological charge 
into a bimeron  
(see Fig.~\ref{fig:chain_num}).
We also have found 
from the effective theory and numerical simulations.\ 
that  
if we rotate the domain wall by applying magnetic fields, 
the domain-wall skyrmion crystal 
becomes wavy as shown in 
Fig.~\ref{fig:chain_effth_nonzero}.

As found in this paper, the domain-wall skyrmions do not exhibit cusps 
for an in-plane easy-axis anisotropy,  
in contrast to the out-of-plane anisotropy for which 
the domain wall has cups at skyrmions.
This fact may have an advantage 
when applying to nanotechnology such as magnetic memories. 
Dynamics of domain-wall skyrmions under applied electric currents should 
be an important direction to explore.

Apart from such applications, there are also some fundamental aspects. 
Namely, 
our paper proposes a novel 
notion of phases of matter, 
{\it i.e.~},
phases with specific boundary conditions 
for instance requiring the presence of topological defects 
(a domain wall in our case). 
That in turn classifies 
phases inside or in the vicinity of 
the topological defects. 
This can be applied to other topological defects, for instance, to vortices.

\begin{acknowledgments}
This work is supported in part by JSPS KAKENHI [Grants No. JP23KJ1881 (YA) and No. JP22H01221, JP23K22492 (MN)], and by the WPI program ``Sustainability with Knotted Chiral Meta Matter (SKCM$^2$)'' at Hiroshima University.
 The numerical computations in this paper were run on the ``GOVORUN" cluster supported by the LIT, JINR.
\end{acknowledgments}


%


\end{document}